\documentclass[preprint,preprintnumbers,amsfonts,amssymb,amsmath,fleqn]{revtex4}
\usepackage{epsf,graphicx}
\usepackage{dcolumn}
\usepackage{bm}
\usepackage{color}
\usepackage{ulem}

\begin{document}

\title{Thermoelectric properties of Bi$_{2}$Te$_{3}$ atomic quintuple thin films}
\bigskip

\author{Ferdows \surname{Zahid}}
\email{fzahid@ee.ucr.edu}
\author{Roger \surname{Lake}}
\email{rlake@ee.ucr.edu}
\affiliation{Department of Electrical Engineering, University of
California, Riverside, CA 92521-0204}

\date{\today}
\medskip
\widetext
\begin{abstract}
Motivated by recent experimental realizations of quintuple atomic layer films of
Bi$_{2}$Te$_{3}$, 
the thermoelectric figure of merit, $ZT$, of the quintuple layer 
is calculated and found to increase by a factor of 10 ($ZT = 7.2$)
compared to that of the bulk at room temperature. 
The large enhancement in $ZT$ results from the change in the distribution of the valence band density of modes
brought about by the quantum confinement in the thin film.
The theoretical model uses \textit{ab initio} electronic structure calculations (VASP)
with full quantum-mechanical structure relaxation combined with
a Landauer formalism for the linear-response transport coefficients.
%
\end{abstract} 

\bigskip
\maketitle

In thermoelectric device applications a high value of the figure of merit ($ZT$) is desirable
for greater efficiency.\cite{rGold,rIoffe} The thermoelectric figure of merit is defined as $ZT=S^{2}\sigma T/\kappa$,
where $T$ is the temperature, $S$ is the Seebeck coefficient, $\sigma$ is the electrical conductivity, and
$\kappa$ is the thermal conductivity which is the sum of the electronic ($\kappa_{e}$) and the lattice ($\kappa_{l}$)
contributions. 
For thermoelectric devices, a $ZT$ value of above 3 or 4 at the room temperature is required to be
competitive with the conventional methods.\cite{rSalvo} 
For the thermoelectric applications Bi$_{2}$Te$_{3}$ has become a material of
particular interest, 
since it gives the highest $ZT$ value of $0.68$ at room temperature in its bulk form.\cite{rGold} 
Recently, interest in this material has been further heightened
by the fact that it is a three dimensional topological insulator \cite{rKane, rHasan}, 
and it has been predicted that one-dimensional topologically protected modes
in line dislocations can significantly increase $ZT$. \cite{ZT_TE_dislocations_APL10}
Many different approaches have been proposed and 
attempted to improve the thermoelectric performance of Bi$_{2}$Te$_{3}$, 
namely, by suppressing the lattice thermal conductivity, \cite{rDress3, rBalan2, rQuinn} 
by tuning carrier concentration, by engineering the band structure,\cite{rHer} and
by reducing the device dimensionality.\cite{rDress1, rDress2} 
The recent reports of the mechanical exfoliation and growth of quintuple layers of
Bi$_{2}$Te$_{3}$ \cite{rBalan1,Balandin_BiTe_APL10} and Bi$_2$Se$_3$ \cite{Crossover_3D_2D_BiSe_MatPhys10} 
motivate this study and 
underscore the importance of developing theoretical models to study the thermoelectric properties of
few-atomic-layer thin films.

The existing theoretical studies on the thermoelectric devices are mostly based
on either the continuum models \cite{rDress1, rDress2, rMahan1, rMahan2, rBalan3} 
or the tight-binding methods.\cite{rAll, rPurdue1} 
Continuum models (effective mass approximation) though useful for their simplicity and computational efficiency, cannot capture
the details of the atomistic effects that can be present in low-dimensional structures.
On the other hand, tight-binding methods usually provide an accurate atomistic description of 
the bandstructures, however, these methods often employ a large number of empirical parameters
obtained by fitting to experimental data or \textit{ab initio} results that are available 
for the \textit{bulk} materials. 
These parameters may not be transferable to low-dimensional structures where surface and interface
issues can become important. 
In this study, we employ a state-of-the-art, density functional theory (DFT), \textit{ab initio} method
to calculate the electronic structures of bulk and two-dimensional (2D) Bi$_{2}$Te$_{3}$. 
The thermoelectric parameters are then derived
from the density of modes $M(E)$ (i.e. a distribution of the conducting channels in energy) obtained from the 
electronic bandstructures.
One advantage of the method is that it can be directly applied to low-dimensional structures.
The calculations
show that a room-temperature $ZT$ value as high as $7.15$ (around $10$-fold increase over the bulk value) can be achieved for
atomically-thin 2D films of Bi$_{2}$Te$_{3}$. 

\begin{center}
\begin{figure}[!tb]
{\epsfxsize 6.5 cm \epsfbox{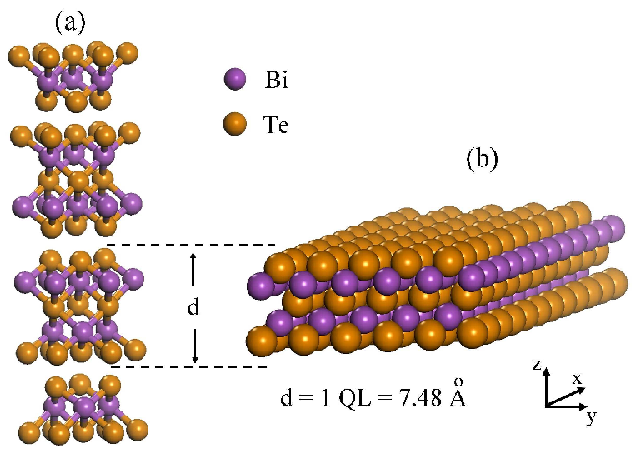}} \vspace{0.1cm}
\caption{}
\label{fig1}
\end{figure}
\end{center}

Fig.~\ref{fig1} shows the atomic structures of the two systems we study in this work:
(a) a rhombohedral crystal structure of the bulk Bi$_{2}$Te$_{3}$ with the space group
D$_{3d}^{5}$ (R$\bar{3}$m), it consists of five-atomic layers (Te-Bi-Te-Bi-Te) arranged
along the $z$-direction, known as quintuple layers (QL). The quintuple layers are separated
from each other by weak van der Waals forces; (b) a free-standing film of 
Bi$_{2}$Te$_{3}$ with the thickness of $1$ QL. 
In this work, all of the \textit{ab initio} calculations (geometry optimizations and 
electronic structures) have been  carried out using a projector augmented wave method within the
framework of the Perdew-Burke-Ernzerhof (PBE)-type generalized gradient approximation of the 
density functional theory as implemented in the software package VASP.\cite{rKresse} 
The Monkhorst-Pack scheme is adopted for the integration
of Brillouin zone with a k mesh of $6 \times 6 \times 6$ for the bulk and $12 \times 12 \times 1$
for the thin film. An energy cutoff of $400$ eV is used in the plane wave basis. The optimized lattice 
parameters for the bulk are: $a = 4.4842$ {\AA} and $c = 31.3767$ {\AA} whereas the
optimized thickness of the $1$ QL thin film is $d = 7.48$ {\AA}.
Spin-orbit coupling is included in the calculations of the electronic structure.

\begin{center}
\begin{figure}[!tb]
{\epsfxsize 6.5 cm \epsfbox{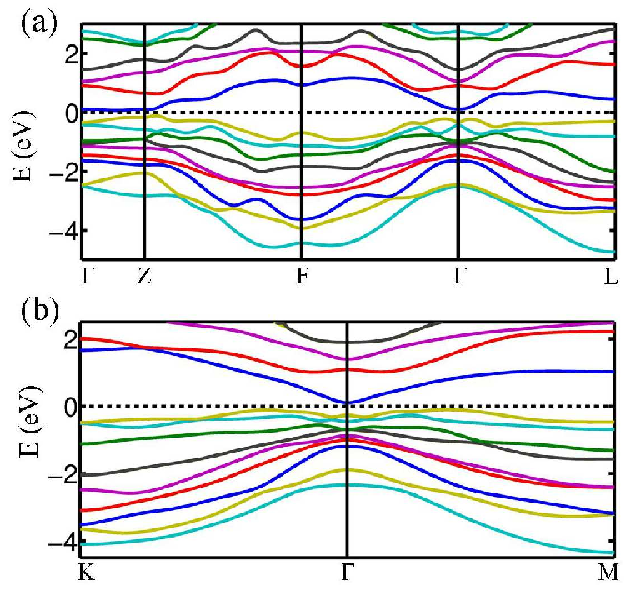}} \vspace{0.1cm}
\caption{} 
\label{fig2}
\end{figure}
\end{center}

Fig.~\ref{fig2} shows the \textit{ab initio} electronic bandstructures of the bulk and QL film.
The results are consistent with previous calculations for both bulk\cite{rMishra} and thin films \cite{rZhang}. 
The calculated bandgap energy ($E_{g}$) of $90$ meV for bulk Bi$_{2}$Te$_{3}$ is 
slightly smaller than the experimental bandgap of $150$ meV. 
For the QL thin film, the bandgap increases
to $190$ meV due to quantum confinement effects. 
The underestimation of the bandgap
energy in semiconductor materials with the DFT methods is a familiar issue. 
In order to achieve
a better agreement with the experimental results we apply the so called `scissors operator'
and adjust the bandgap of the bulk Bi$_{2}$Te$_{3}$ to match the experimental value
by rigidly shifting the conduction bands up and the valence bands down by $30$ meV around the midgap energy.
The bandgap of the thin film is adjusted by the same amount. 
In this way, the calculated
increase in the bandgap energy of the QL film compared to that of the bulk is left unchanged.
After the rigid shift of the bands the bandgap energies become $E_G = 150$ meV for the 
bulk and $E_G = 250$ meV for the QL film.
For the calculations of thermoelectric properties, we set the zero in the energy scale at the midgap energy 
for both the bulk and the QL film.
%


The next step is to derive the thermoelectric (TE) parameters from the \textit{ab initio} 
bandstructures. The TE parameters are usually evaluated from
the Boltzmann transport equation (BTE).\cite{rGold} 
An alternative approach is the Landauer formalism\cite{rLand, rDatta1} which
is more convenient for performance comparisons between materials of different dimensions. \cite{rPurdue2}
In this study, the objective is to compare the TE parameters between the 3D bulk and the 2D film, hence we have chosen
the Landauer approach. Within the Landauer formalism in the linear response regime, the
electronic conductivity ($\sigma$), thermal conductivity for zero electric current ($\kappa_{e}$),
and the Seebeck coefficient (S) are expressed as \cite{rPurdue1}

\begin{eqnarray}
\sigma &=& (2q^{2}/h)I_{0}\quad [\mathrm{1/\Omega-m}], \nonumber \\
\kappa_{e} &=& (2Tk_{B}^{2}/h)(I_{2} - I_{1}^{2}/I_{0}) \quad [\mathrm{W/K-m}], \nonumber \\
 S &=& -(k_{B}/q)\frac{I_{1}}{I_{0}}\quad [\mathrm{V/K}],\nonumber \\ 
\mathrm{with} \nonumber \\
I_{j} &=& L \int_{-\infty}^{\infty} \left(\frac{E-E_{F}}{k_{B}T}\right)^{j} \bar{T}(E)\left(-\frac{\partial f_{0}}{\partial E}\right)dE  
\label{eq.main}
\end{eqnarray}

\noindent
where $L$ is the device length and the transmission function
$\bar{T}(E) = T(E)M(E)$ with M(E) as the density of modes (DOM). \cite{rDatta1} 
In the diffusive limit,
$T(E)=\lambda(E)/L$ with $\lambda(E)$ as the electron mean free path. \cite{rDatta2} 
When phonon scattering is dominant, 
the mean free path can be written as $\lambda(E)=\lambda_{0}$, a constant. \cite{rPurdue1} 
The density of modes M(E)
can be expressed as \cite{rDatta2, rPurdue1}

\begin{equation}
M(E)=\sum_{k_{\perp}} \Theta(E-\epsilon({k_{\perp}}))
\label{eq.ME}
\end{equation}

\noindent
where $\Theta$ is the unit step function, and $k_{\perp}$ refers to all the $k$ states in the
first Brillouin zone perpendicular to the transport direction. 
Using Eq. (\ref{eq.ME}), the DOM in
any dimension can be numerically evaluated from a given $\epsilon(k)$ simply by counting the bands that cross
the energy of interest. Note that the expressions and the units indicated in Eq.(\ref{eq.main}) are same in all three 
dimensions provided the DOM is expressed in per unit area in each dimension.

\begin{center}
\begin{figure}[!tb]
{\epsfxsize 5.5 cm \epsfbox{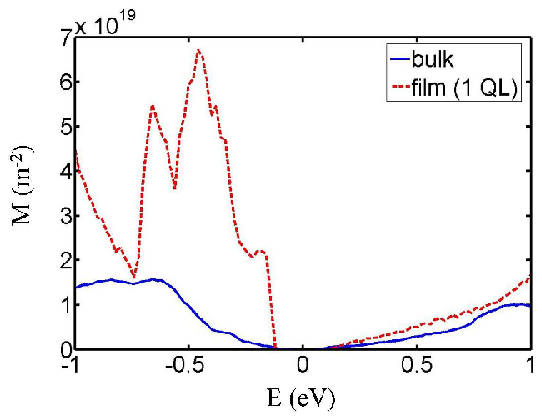}} \vspace{0.1cm}
\caption{} 
\label{fig3}
\end{figure}
\end{center}

Fig.~\ref{fig3} shows the density of modes calculated from the \textit{ab initio} bandstructures using
Eq.~(\ref{eq.ME}). 
For the integration over the first Brillouin zone, k points 
are sampled on a uniform rectangular grid. 
The convergence of the final results 
has been ensured by using sufficient number of k points ($101 \times 51 \times 10$ for the bulk
and $151 \times 101 \times 1$ for the thin film).
For the conduction band ($E>0$), 
a slight increase in the value of the $M(E)$ is observed for the QL film 
while the shape remains almost same as in the bulk. 
However, for the valence band ($E<0$), the value of $M(E)$ for the QL is much higher compared to that
of the bulk. 
Most importantly, at the valence band edge, the DOM for the QL turns on abruptly and has a peaked distribution.
Previous work showed that a delta-shaped transport distribution function of the BTE 
(equivalent to M(E) in the Landauer formalism \cite{rPurdue1})
maximized the $ZT$. \cite{rMahan1}

\begin{center}
\begin{figure}[!tb]
{\epsfxsize 6.5 cm \epsfbox{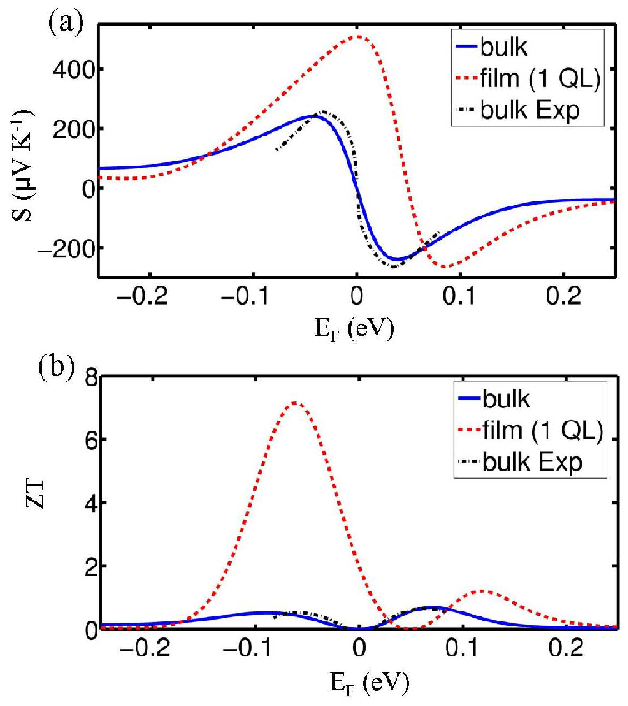}} \vspace{0.1cm}
\caption{}
\label{fig4}
\end{figure}
\end{center}

Thermoelectric parameters for the bulk and the QL film are calculated at the room temperature 
from Eq.(\ref{eq.main})
using the density of modes presented above.
The results for the Seebeck coefficient ($S$) and the figure of merit 
($ZT$) are shown in Fig.~\ref{fig4}. 
Values of the electron and hole mean free paths of
$\lambda_{0} = 14$ nm (for the conduction band)
and  $\lambda_{0} = 8$ nm (for the valence band) give the best agreement with the bulk experimental data, 
and they are consistent with those used in previous studies.\cite{rAll,rPurdue1} 
For the lattice thermal conductivity $\kappa_{l}$
we use the experimental value of 1.5 WK$^{-1}$m$^{-1}$. 
The results for the bulk
Bi$_{2}$Te$_{3}$ show excellent agreement with the experimental data. 
For the thin film, there is a large increase
in the thermoelectric parameters. 
The maximum in the Seebeck coefficient ($S$) increases by more than two times that of 
the bulk value, and the maximum $ZT$ value of $7.15$ obtained at $E_{F}=-0.06$ eV is around ten times
higher than the bulk value of $0.68$. 
The enhancements in the TE parameters are due to the improvements
in the magnitude and the shape of the density of modes of the thin film brought about by the 
confinement in 2D. 
Note that for the thin film we use the same values for the parameters 
$\lambda_{0}$ and $\kappa_{l}$ as in the bulk. 
%
In reality, surface roughness, defects, and interface scattering present
in 2D thin films can reduce those values.

In conclusion, \textit{ab initio} electronic structure calculations combined with a Landauer
approach for the linear-response transport coefficients
show that the thermoelectric properties of the recently 
obtained Bi$_2$Te$_3$ quintuple layers can be significantly enhanced from those of the bulk.


\textbf{Acknowledgements.} This work is supported by the Microelectronics Advanced Research
Corporation Focus Center on Nano Materials (FENA). F.Z. would like to 
thank Prof. Hong Guo and RQCHP for providing the computational resources.

\newpage
\noindent
{\bf \center Figure captions} \\
\\
\noindent Fig.~\ref{fig1}. (color online) Atomic structures of Bi$_{2}$Te$_{3}$:
(a) bulk unit cell; (b) free-standing slab with thickness 
of one quintuple layer (QL). Transport is assumed to be in the direction of the $x$ axis (binary axis).
\\
\\
\noindent Fig.~\ref{fig2}.(color online) \textit{Ab initio} bandstructures of Bi$_{2}$Te$_{3}$:
(a) bulk; (b) thin film of $1$ QL thickness. The broken lines
shows the position of the Fermi energy. For the bulk the symmetry points
are given by: $\Gamma(0,0,0)$, $F(-\frac{1}{2},0,-\frac{1}{2})$, $L(0,\frac{1}{2},0)$, and
$Z(\frac{1}{2}, \frac{1}{2}, \frac{1}{2})$. 
For the thin film, the points are:
$\Gamma(0,0)$, $K(\frac{1}{3},\frac{1}{3})$, and $M(\frac{1}{2},0)$. All the points are 
in the unit of $(2\pi/a)$ in the reciprocal space.
\\
\\
\noindent Fig.~\ref{fig3}. (color online) Density of modes M(E) per unit area for Bi$_{2}$Te$_{3}$: solid line 
for bulk and the broken line for $1$ QL thick thin film. The midgap energy is set at $E = 0$. 
Note the sharp increase in M(E) for the thin film just below the midgap energy in the valence band.
\\
\\
\noindent Fig.~\ref{fig4}. (color online) (a) The Seebeck coefficient ($S$) and (b) the figure of merit ($ZT$) 
at the room temperature ($300$K) as a function of the Fermi energy $E_{F}$ for
bulk and thin film of $1$ QL thickness. 
The dotted black line shows the experimental data for bulk.
The electron and hole mean free paths ($\lambda_{0}$) are $14$ nm
and $8$ nm, respectively, which are obtained by fitting our results to the bulk experimental data. 
The zero in the energy scale represents the midgap energy.
\\
\\

\newpage

\newpage

\newpage

\newpage

\end{document}